\documentclass[conference,romanappendices,10pt]{IEEEtran}	

\IEEEoverridecommandlockouts

\usepackage{amssymb,amsmath,amsthm}
\usepackage{mathrsfs}
\usepackage[utf8]{inputenc}
\usepackage[T1]{fontenc}
\usepackage{graphicx}
\usepackage{epsfig}
\usepackage[USenglish]{babel}
\usepackage{subfigure}
\usepackage{color}
\usepackage{import}
\usepackage{cite}
\usepackage{epstopdf}
\usepackage{balance}

{		\newtheorem{theorem}{Theorem}}
{ 		}
{ 		}
{ 		}
{ 		\newtheorem{remark}{Remark}}
{		}
{ 		}

\newcommand{\h}{\mathbf{h}}

\newcommand{\n}{\mathbf{n}}

\renewcommand{\v}{\mathbf{v}}
\newcommand{\w}{\mathbf{w}}

\renewcommand{\H}{\mathbf{H}}
\newcommand{\I}{\mathbf{I}}

\newcommand{\W}{\mathbf{W}}

\newcommand{\setC}{\mathcal{C}}

\newcommand{\setN}{\mathcal{N}}

\newcommand{\Compl}{\mbox{$\mathbb{C}$}}

\newcommand{\Exp}{\mathbb{E}}
\newcommand{\herm}{\mathrm{H}}

\newcommand{\tr}{\mathrm{tr}}

\begin{document}

\title{Optimal Low-Complexity Self-Interference \\ Cancellation for Full-Duplex MIMO Small Cells}

		
\author{\IEEEauthorblockN{Italo Atzeni, Marco Maso, and Marios Kountouris}%
		\IEEEauthorblockA{Mathematical and Algorithmic Sciences Lab, France Research Center, Huawei Technologies Co. Ltd. \\
		\{italo.atzeni, marco.maso, marios.kountouris\}@huawei.com}}

\maketitle

\begin{abstract}
Self-interference (SI) significantly limits the performance of full-duplex (FD) radio devices if not properly cancelled. State-of-the-art SI cancellation (SIC) techniques at the receive chain implicitly set an upper bound on the transmit power of the device. This paper starts from this observation and proposes a transmit beamforming design for FD multiple-antenna radios that: i) leverages the inherent SIC capabilities at the receiver and the channel state information; and ii) exploits the potential of multiple antennas in terms of spatial SIC. The proposed solution not only maximizes the throughput while complying with the SIC requirements of the FD device, but also enjoys a very low complexity that allows it to outperform state-of-the-art counterparts in terms of processing time and power requirements. Numerical results show that our transmit beamforming design achieves significant gains with respect to applying zero-forcing to the SI channel when the number of transmit antennas is small to moderate, which makes it particularly appealing for FD small-cell base stations.
\end{abstract}


\section{Introduction} \label{sec:intro}
Full-duplex (FD) radio has recently developed from long-studied theoretical concept to potential candidate solution to increase the performance of future wireless networks. By transmitting and receiving simultaneously on the same frequency band, FD radios can theoretically double the throughput with respect to their half-duplex (HD) counterparts \cite{Sab14}. Despite the potential of the FD approach, a critical issue hinders the effective achievement of the promised performance gains: the transmit and receive antennas of the FD radio need to be perfectly isolated, although physical limitations do not allow to attain this condition. As a result, strong self-interference (SI) appears at the receive chain, with consequent reduction of the signal-to-interference-plus-noise ratio (SINR) of the received signal. This issue implicitly sets an upper bound on the transmit power of the device. In this respect, small-cell base stations (BSs) prove especially suitable for the deployment of FD technology thanks to their low transmit power and the low mobility of the user terminals (UTs) \cite{Atz15a}.

Several efforts have been devoted in recent years to design effective SI cancellation (SIC) techniques as a means to approach the theoretical throughput of FD communications. In general, perfect SIC is assumed to be achieved whenever the SI is reduced to the same level as the noise floor. In this context, the best SIC results so far, for both single- and multiple-antenna FD settings, are accomplished by means of hybrid solutions based on both analog and digital signal processing~\cite{Bha13,Bha14}:
\begin{itemize}
	\item[i)] Performing part of the cancellation via analog signal processing is beneficial to reduce problems such as saturation of the amplifiers and low dynamic range at the analog-to-digital converter of the receive chain~\cite{Knox12, Maso15FD};
	\item[ii)] By resorting to digital signal processing at both transmit and receive side, one can exploit additional degrees of freedom and support higher transmit power of the FD device while preserving the effectiveness of SIC \cite{Ahm15, Kor14}.
\end{itemize}
In this regard, it is worth noting that the potential of digital signal processing for SIC is larger in case of multiple-input multiple-output (MIMO) systems, for which the advantages in terms of additional degrees of freedom can be exploited. The solutions based on this approach are typically referred to as \textit{spatial SIC} strategies.

As a matter of fact, a fundamental tradeoff exists between the effectiveness of spatial SIC and the achievable throughput. In practice, the more transmit (resp. receive) antennas are devoted to suppressing the SI, the less power is conveyed to (resp. from) the desired link. In addition, it is generally more convenient to perform such spatial SIC at the transmitter than at the receiver for a two-fold reason:
\begin{itemize}
	\item[i)] The transmit power of the FD device is consistently higher than the power of the desired incoming signal and it is thus meaningful to fully exploit the receive antennas to maximize the signal-to-noise ratio (SNR) of the latter;
	\item[ii)] An overabundance of SI may saturate the receiver circuitry, preventing any spatial SIC at the receive chain to be applied in the first place.
\end{itemize}
A possible approach is to design the transmit beamformer so as to apply full zero-forcing (ZF) to the SI channel, which allows to null the SI entirely in case of perfect channel state information (CSI) \cite{Rii11}. However, this solution does not exploit the aforementioned analog/digital SI capabilities at the receive chain: in fact, since the FD device can tolerate the SI power to be up to a certain threshold, nulling the SI completely via spatial SIC proves excessively aggressive and results in lower throughput. In this regard, an optimal transmit beamforming design in which the throughput of the downlink transmission is maximized under SI constraints was proposed in \cite{Zha12}. Therein, an iterative search algorithm based on well-known convex optimization techniques is proposed to identify the optimal transmit beamformer. Nevertheless, such scheme may not be adequate to compute the optimal solution within the coherence time of the wireless channel in realistic settings, especially in rich scattering environments populated by UTs whose channels can experience rapid variations in their fast-fading components.

Starting from these observations, in this paper we consider an FD MIMO radio with partial SIC capabilities at the receive chain: assuming perfect CSI of both the downlink and the SI channel, we propose a transmit beamforming design that leverages:
\begin{itemize}
	\item[a)] The inherent SIC capability of the device;
	\item[b)] The potential of multiple antennas in terms of spatial SIC.
\end{itemize}
Remarkably, the proposed solution is not only optimal in terms of throughput, but also enjoys a very low complexity that allows it to outperform state-of-the-art methods in terms of processing time and power requirements. In fact, the resulting transmit beamformer is characterized by a simple closed-form expression that follows from a practically relevant optimization problem formulation, whereas existing approaches for multiple-antenna FD radios achieve the desired SIC only at the expense of higher complexity, i.e., through more complex iterative algorithms (e.g. \cite{Zha12}). Evidently, this one-shot solution results in a complexity gain that is increasingly appealing as the number of antennas at the BS grows: in fact, the higher the number of transmit antennas at the BS, the larger the search set for iterative algorithms such as in \cite{Zha12}. Hence, the spatial SIC cancellation presented in this paper outperforms state-of-the-art counterparts in terms of both performance (cf. \cite{Rii11}) and complexity (cf. \cite{Zha12}). More specifically, numerical results show that the proposed approach achieves significant gains when the number of transmit antennas is small to moderate, which makes it particularly suitable for implementation in FD small-cell BSs.

\section{System Model} \label{sec:SM}

\begin{figure}[t!]
	\centering
	\includegraphics[scale=1]{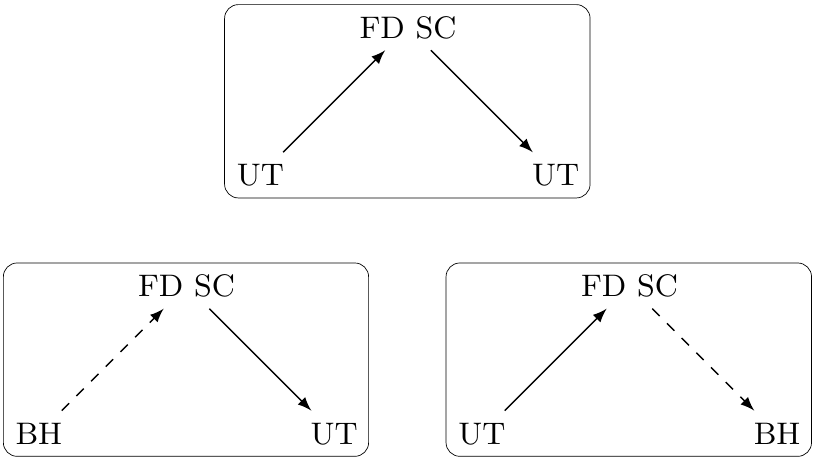} 
	\caption{Possible scenarios for the operation of FD small-cell (SC) BSs; solid and dashed lines indicate short- and long-range links, respectively.} \label{fig:conf} 
\end{figure}

Consider a hybrid FD/HD scenario where, at each timeslot, a multiple-antenna FD BS serves one HD node in the uplink and one HD node in the downlink, both single-antenna. Figure~\ref{fig:conf} depicts three possible instances of such hybrid FD/HD scenario in the context of small-cell networks, namely:
\begin{itemize}
\item[i)] UTs in both uplink and downlink;
\item[ii)] Backhaul (BH) BS in the uplink and UT in the downlink;
\item[iii)] UT in the uplink and BH BS in the downlink.
\end{itemize}
Two observations are in order at this stage. The hybrid FD/HD network configuration, i.e., a scenario where FD and HD devices operate side by side, is arguably the most suitable setting for our study, since it avoids strong inter-node interference while exploiting the full throughput gain provided by the FD paradigm at the BSs~\cite{Goyal15}. Additionally, the single-user communication assumed to occur in the uplink/downlink does not diminish the generality of our approach at the physical layer: in fact, this could be seen as the result of scheduling decisions, typically performed at upper layers in current networks. For clarity of presentation, in the rest of the paper we assume scenario (i).

Now, let $N_{R}$ and $N_{T}$ be the number of receive and transmit antennas, respectively. The task of the FD BS in this context is to exploit (all or a subset of) its receive antennas to maximize the uplink throughput and, simultaneously, to exploit (all or a subset of) its transmit antennas to maximize the downlink throughput. Moreover, as previously discussed, it is assumed that the FD BS has preexisting hybrid SIC capabilities, such as the ones described in Section~\ref{sec:intro}. In this context, the FD BS needs to guarantee that the power of the SI experienced during the reception of the incoming signal does not exceed a certain threshold, which guarantees the full effectiveness of the preexisting hybrid SIC capabilities at the receive chain. It is worth noting that the latter are typically characterized in the literature in terms of maximum SI attenuation/cancellation they can provide, expressed in dB~\cite{Bha13,Bha14}. In practice, each of these strategies sets an implicit upper bound on the maximum transmit power that can be adopted by the FD radio in order to preserve the full effectiveness of the preexisting SIC algorithms (and the throughput of the incoming transmission). In particular, this can be straightforwardly obtained by summing the maximum SIC capability to the noise floor and by subsequently subtracting inter-antenna distance-dependent pathloss attenuation experienced by the transmitted signal during its propagation from the transmit to the receive antennas.

Concerning the notation adopted throughout the paper, it is convenient to begin by clearly differentiating between uplink 
(i.e., from the UT served in the uplink to the FD BS) and downlink communication (i.e., from the FD BS to the UT served in the downlink). In this context, we let $\h_{\mathrm{u}} \in \Compl^{N_{R}}$ and $\h_{\mathrm{d}} \in \Compl^{N_{T}}$ be the uplink and the downlink channels, respectively. Likewise, we define $p_{\mathrm{u}}$ and $p_{\mathrm{d}}$ as the uplink and downlink transmit powers, respectively. We use $s_{\mathrm{u}}$ and $s_{\mathrm{d}}$ to denote the uplink and downlink data symbols, respectively, with $\Exp [|s_{\mathrm{u}}|^{2}] = 1$ and $\Exp [|s_{\mathrm{d}}|^{2}] = 1$. Furthermore, we let $\n_{\mathrm{u}} \sim \setC \setN(0,\sigma^{2} \I_{N_{R}})$ and $n_{\mathrm{d}} \sim \setC \setN(0,\sigma^{2})$ be the additive noise in the uplink and in the downlink, respectively. Finally, we let $\H \in \Compl^{N_{R} \times N_{T}}$ be the SI channel at the FD BS and denote the receive combiner and the transmit beamformer used by the FD BS with the vectors $\v \in \Compl^{N_{R}}$ and $\w \in \Compl^{N_{T}}$, respectively.

We assume that $\h_{\mathrm{d}}$ is subject to Rayleigh fading with elements distributed independently as $\setC \setN (0, 1)$, whereas $\H$ is subject to Ricean fading \cite{Dua12} and, therefore, its elements are distributed independently as $\setC \setN (\mu, \nu^{2})$. In this regard, one can measure the Ricean $K$-factor and the pathloss $\Omega$ between transmit and receive antennas and determine the mean and standard deviation of $\H$ as (cf. \cite{Atz15a})
\begin{align*} 
\mu \triangleq \sqrt{\frac{K \Omega}{K+1}}, \qquad \nu \triangleq \sqrt{\frac{\Omega}{K+1}}.
\end{align*}
We assume that the FD BS has perfect CSI related to both $\h_{\mathrm{d}}$ and $\H$. The uplink and downlink signals can be expressed as
\begin{align*}
y_{\mathrm{u}} & \triangleq \sqrt{p_{\mathrm{u}}} \v^{\herm} \h_{\mathrm{u}} s_{\mathrm{u}} + \sqrt{p_{\mathrm{d}}} \v^{\herm} \H \w s_{\mathrm{d}} + \v^{\herm} \n_{\mathrm{u}} \\
y_{\mathrm{d}} & \triangleq \sqrt{p_{\mathrm{d}}} \h_{\mathrm{d}}^{\herm} \w s_{\mathrm{d}} + n_{\mathrm{d}},
\end{align*}
respectively. Lastly, we let $\varepsilon > 0$ be the SI threshold, i.e., the maximum tolerable power of the SI experienced at the receive antennas to preserve the full effectiveness of the pre-existing SIC algorithms. In a practical case, this is given by
\begin{align} \label{eq:epsilon}
\varepsilon = r_{\mathrm{n}} - c
\end{align}
where $r_{\mathrm{n}}$ represents the noise floor and $c$ is the SIC capability at the receive chain.

\section{Transmit Beamforming Design} \label{sec:TBD}


Similarly to \cite{Zha12}, we aim at maximizing the downlink spectral efficiency, i.e., from the FD BS to the served UT, while keeping the SI below a certain threshold. The resulting optimization problem can be written as follows:
\begin{align} \label{eq:opt1} \tag{$\mathrm{P}$}
\begin{array}{ccll} \vspace{2mm}
\displaystyle \max_{\w} & & \log_{2} \big( 1 + \rho |\h_{\mathrm{d}}^{\herm} \w|^{2} \big) & \\
\displaystyle \mathrm{s.t.} & & |\v^{\herm} \H \w|^{2} \leq \varepsilon \\
\displaystyle & & \| \w \|^{2} \leq 1
\end{array}
\end{align}
where $\rho \triangleq p_{\mathrm{d}}/\sigma^{2}$ is the SNR at the downlink UT. Furthermore, switching our focus to the constraints in \eqref{eq:opt1}, we note that the first constraint enforces that the SI experienced at the receive antennas does not exceed the threshold that guarantees the full effectiveness of the pre-existing SIC algorithms,\footnote{Herein, we assume that the receive combining vector $\v$ is independent from the SI channel as in, e.g., maximum ratio combining.} whereas the second constraint ensures that the solution does not induce any undesired amplification of the transmit signal.

Now, let $(\cdot)^{\sharp}$ denote the Moore-Penrose pseudoinverse operator and let $\I$ be the $N_{T}$-dimensional identity matrix. In the next theorem, we provide the closed-form expression of the optimal transmit beamformer for the considered problem.

\begin{theorem} \rm{
The transmit beamformer that solves \eqref{eq:opt1} is given by
\begin{align} \label{eq:w_star}
\w^{\star} & \triangleq \frac{( \I - \alpha^{\star} \H^{\herm} \v (\H^{\herm} \v)^{\sharp}) \h_{\mathrm{d}}}{\| ( \I - \alpha^{\star} \H^{\herm} \v (\H^{\herm} \v)^{\sharp}) \h_{\mathrm{d}} \|}
\end{align}
with
\begin{align}
\label{eq:alpha_star} \alpha^{\star} & \triangleq 1 - \min \bigg( 1, \sqrt{\frac{\max(0, \zeta - \eta)}{\zeta}} \bigg) 
\end{align}
where we have defined
\begin{align}
\label{eq:zeta} \zeta \triangleq \ & \bigg( 1 - \frac{\varepsilon}{\v^{\herm} \H \H^{\herm} \v} \bigg) \h_{\mathrm{d}}^{\herm} \H^{\herm} \v \v^{\herm} \H \h_{\mathrm{d}} \\
\label{eq:eta} \eta \triangleq \ & \h_{\mathrm{d}}^{\herm} \H^{\herm} \v \v^{\herm} \H \h_{\mathrm{d}} - \varepsilon \h_{\mathrm{d}}^{\herm} \h_{\mathrm{d}}.
\end{align}}
\end{theorem}

\begin{figure}[!t]
	\centering
	{\def\svgwidth{\columnwidth}
		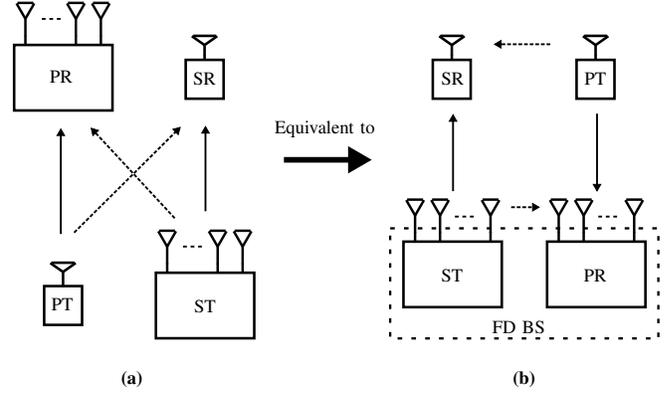}
	\caption{Virtual model conversion from a cognitive interference channel with interference temperature constraint (a) to an SI FD channel (b).}
	\label{fig:conversion} 
\end{figure}

\begin{IEEEproof}
The key enabler to derive a closed-form expression of the optimal transmit beamformer is the intuition that the considered FD SI channel is equivalent to a cognitive interference channel with an interference temperature constraint. In this respect, Figure~\ref{fig:conversion}(a) depicts a licensed system consisting of a primary transmitter/receiver pair (PT/PR) that coexists with an opportunistic system consisting of a secondary transmitter/receiver pair (ST/SR), both operating in the same bandwidth. Let us assume an interference temperature constraint imposed on the ST, i.e., the maximum interference that the opportunistic transmission can generate towards the PR is upper bounded. Now, the equivalence between this setting and the FD SI channel considered in this paper becomes evident from Figure~\ref{fig:conversion}(b), where the same labels as in Figure~\ref{fig:conversion}(a) are used for convenience. In particular, we note that:
\begin{itemize}
    \item[-] The transmit antennas of the FD BS serve a UT in the downlink, and such transmission causes SI at the receive antenna array, which is simultaneously communicating with another device in the uplink.
	\item[-] In order to guarantee the full effectiveness of the SIC algorithms at the receive chain, an upper bound on the maximum tolerable power of the SI is imposed. Such upper bound has the same role as the interference temperature constraint in the aforementioned cognitive interference channel.
\end{itemize}
Thus, it is not difficult to see that the receive and transmit antennas of the FD BS are equivalent to a virtual PR and ST, respectively. Similarly, the UT served in the downlink operates as virtual SR, whereas the device communicating in the uplink acts as virtual PT. Therefore, we can cast \eqref{eq:opt1} into a rate maximization problem for a cognitive interference channel with interference temperature constraint, whose solution is known to have the form \cite{Lv12}
%
%
%
%
%
%
\begin{align} \label{eq:w_alpha}
\w(\alpha) & \triangleq \frac{\alpha \w_{\mathrm{ZF}} + (1 - \alpha) \w_{\mathrm{MRT}}}{\| \alpha \w_{\mathrm{ZF}} + (1 - \alpha) \w_{\mathrm{MRT}} \|}
\end{align}
where we have defined (see Remark~\ref{rem:pzf})
\begin{align}
\label{eq:w_pzf} \w_{\mathrm{ZF}} & \triangleq (\I - \H^{\herm} \v (\H^{\herm} \v)^{\sharp}) \h_{\mathrm{d}} \\
\w_{\mathrm{MRT}} & \triangleq \h_{\mathrm{d}}. \nonumber
\end{align}
In other words, the optimal solution of \eqref{eq:opt1} can be expressed as the (normalized) linear combination of a maximum ratio transmission (MRT) beamformer, obtained as a function of the downlink channel, and a ZF beamformer, obtained as a function of both the SI and the downlink channel. In particular, we note that the normalization constraint is satisfied by construction of $\w(\alpha)$.

Now, it is easy to observe that both the objective and the SI constraint in \eqref{eq:opt1} are monotonically decreasing functions of $\alpha$. Then, it follows that the optimal solution of \eqref{eq:opt1} satisfies the SI constraint with equality. In this regard, let $\widetilde{\alpha}$ be the (unbounded) solution of $|\v^{\herm} \H \w(\alpha)|^{2} = \varepsilon$ given by
\begin{align} \label{eq:alpha_tilde}
\widetilde{\alpha} \triangleq 1 - \sqrt{\frac{\zeta - \eta}{\zeta}}
\end{align}
with $\zeta$ and $\eta$ defined in \eqref{eq:zeta} and \eqref{eq:eta}, respectively; then, we readily obtain $\alpha^{\star}$ in \eqref{eq:alpha_star} from \eqref{eq:alpha_tilde} by imposing $\widetilde{\alpha} \in [0,1]$. Finally, $\w^{\star}$ is computed as in \eqref{eq:w_star}.
\end{IEEEproof}

\begin{remark} \label{rem:pzf} \rm{
The transmit beamformer $\w_{\mathrm{ZF}}$ in \eqref{eq:w_pzf} is chosen as the (non-normalized) projection of $\h_{\mathrm{d}}$ onto the null space of $\H^{\herm} \v$, i.e., one antenna nulls the SI entirely whereas the remaining $N_{T} - 1$ antennas are used to maximize the power signal to the desired link. On the other hand, with $\w_{\mathrm{MRT}}$, all $N_{T}$ antennas are used to maximize the power signal to the desired link. Therefore, increasing $\alpha$ in \eqref{eq:w_star} (resp. \eqref{eq:w_alpha}) corresponds to applying more SIC and results in less power conveyed to the desired link.}
\end{remark}

\begin{remark} \label{rem:low_compl} \rm{
The model conversion from \eqref{eq:opt1} to its cognitive radio-based interpretation casts a $N_{T}$-dimensional problem into a one-dimensional problem, for which we are able to derive the optimal solution in closed-form. On the one hand, such closed-form expression does not require any iterative algorithm; instead, it consists in a single operation with no matrix inversion whatsoever,\footnote{Observe that $\v^{\herm} \H \H^{\herm} \v$ is a scalar value and, therefore, there is no matrix inversion involved either in \eqref{eq:w_star}, i.e., in the computation of $(\H^{\herm} \v)^{\sharp}$, or in \eqref{eq:zeta}.} which enables the computation of the optimal transmit beamformer within the limited coherence time allowed by the fluctuations of the wireless channel. On the other hand, the model conversion brings full scalability with the number of transmit antennas $N_{T}$ in terms of computational complexity, since it only requires the closed-form computation of the one-dimensional optimal variable $\alpha^{\star}$. In that sense, the advantages brought over state-of-the-art solutions for optimal transmit beamforming design proposed in \cite{Zha12} are evident.}
\end{remark}

\begin{figure}[t!]
	\centering
	\includegraphics[scale=1]{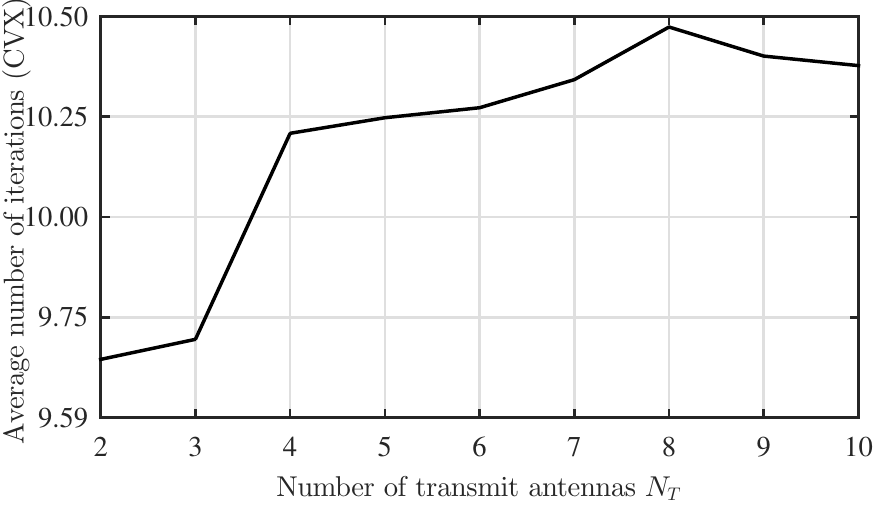} \\ \vspace{2mm}
	\includegraphics[scale=1]{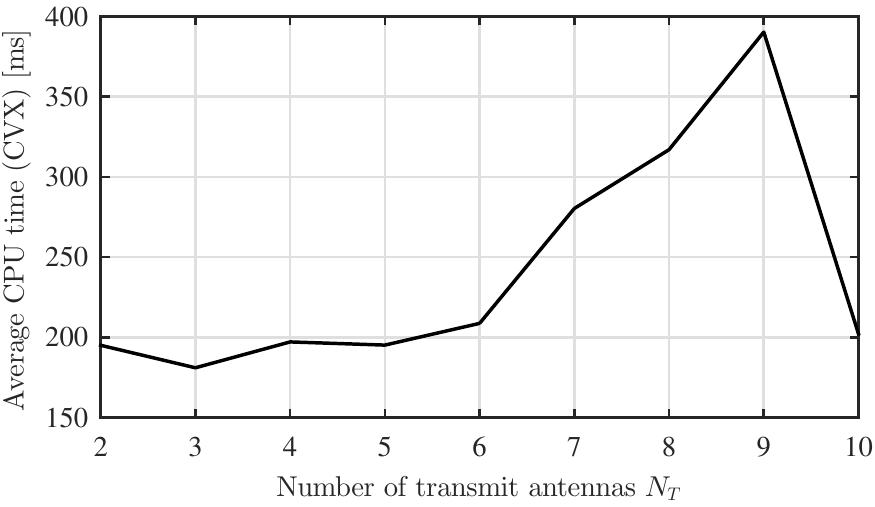} 
	\caption{Average number of iterations and average CPU time using CVX for the computation of the optimal transmit beamformer, with $c=-110$~dB, $\rho=0$~dB, and for different number of transmit antennas.} \label{fig:complexity} \vspace{-1mm}
\end{figure}

\section{Numerical Results} \label{sec:Num}

In this section, we analyze the benefits of the proposed transmit beamforming design with respect to: i) \cite{Zha12} in terms of complexity, and ii) \cite{Rii11} in terms of performance.  The reported numerical results are obtained by means of Monte Carlo simulations over $10^{4}$ channel realizations. In particular, all simulations are run on a single core of a 2.50~GHz CPU with 8~GB of memory. The transmit power and the noise floor of the FD BS are set in compliance with the Long-Term Evolution (LTE) radio frequency planning for outdoor small cells \cite{Holma10}, i.e., $p_{\mathrm{d}} = 30$~dBm and $r_{\mathrm{n}} = -116.4$~dBm. Furthermore, we assume $\Omega=-30$~dB, and $c=-110$~dB as in \cite{Bha13,Bha14}, and that maximum ratio combining (MRC) is adopted at the receive chain of the FD BS, i.e., $\v = \h_{\mathrm{u}}$.

\vspace{2mm}

\noindent{\textbf{Complexity gains.}} We first illustrate the reduced complexity of our approach with respect to \cite{Zha12}. Therein, \eqref{eq:opt1} is solved by applying the following SDP relaxation:
\begin{align} \label{eq:opt3}
\begin{array}{ccl} \vspace{2mm}
\displaystyle \max_{\W} & & \log_{2} \big( 1 + \tr(\h_{\mathrm{d}}^{\herm} \W \h_{\mathrm{d}}) \big) \\
\displaystyle \mathrm{s.t.} & & \W \succeq 0 \\
& & \tr(\v^{\herm} \H \W \H^{\herm} \v) \leq \varepsilon \\
& & \tr(\W) \leq 1
\end{array}
\end{align}
with $\W \triangleq \w \w^{\herm}$. We begin our comparison by solving~\eqref{eq:opt3} numerically with the Matlab-based convex optimization solver CVX \cite{cvx14} and study the complexity of this approach in terms of number of iterations and CPU time. From Figure~\ref{fig:complexity}, we observe that, as could have been intuitively expected, the computation requires an increasing number of iterations and CPU time as the number of antennas increases. In particular, even for low number of antennas, i.e., $N_{T} \in [2,6]$, the CPU time ranges around 200~ms; this is not a suitable value for accommodating the requirements of real-world cellular system implementations, typically characterized by channel coherence times whose largest value is in the order of a few hundreds of ms, even if very low mobility settings are considered \cite{Ghosh10}. In other words, by the time the solution of \eqref{eq:opt1} is found by the algorithm in \cite{Zha12}, the CSI may be outdated and may require a substantial update, de facto rendering the optimal solution useless. Remarkably, this is not the case for our low-complexity optimal transmit beamforming design based on the closed-form expression \eqref{eq:w_star}. In fact, the latter does not require any iterative algorithm to be computed and allows to obtain $\w^{\star}$ with one-shot computation (see Remark~\ref{rem:low_compl}).

\begin{figure}[t!]
\centering
\includegraphics[scale=1]{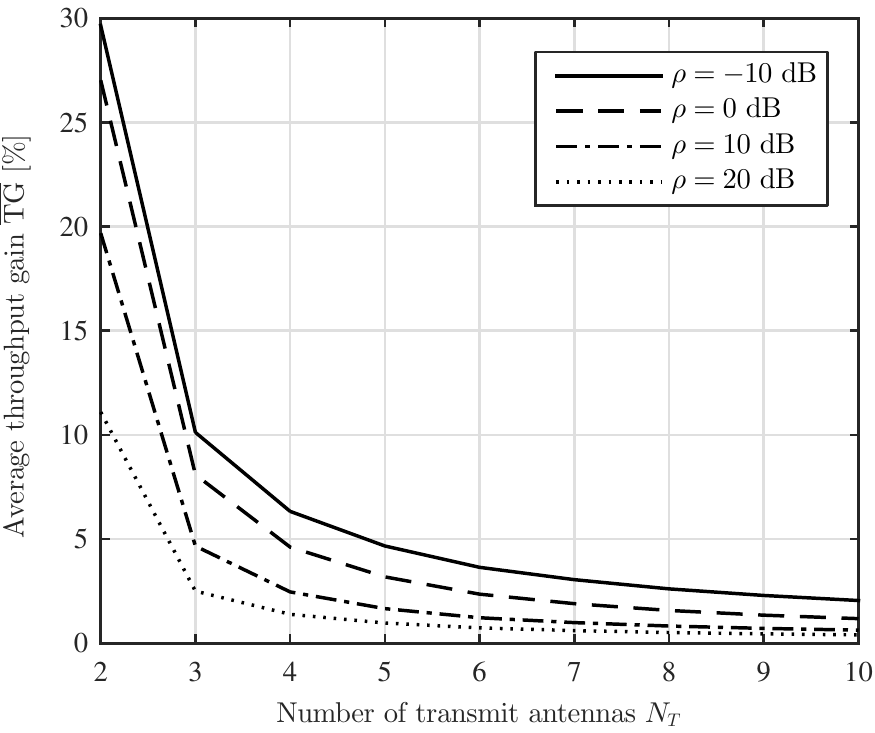} \\ \vspace{2mm}
\includegraphics[scale=1]{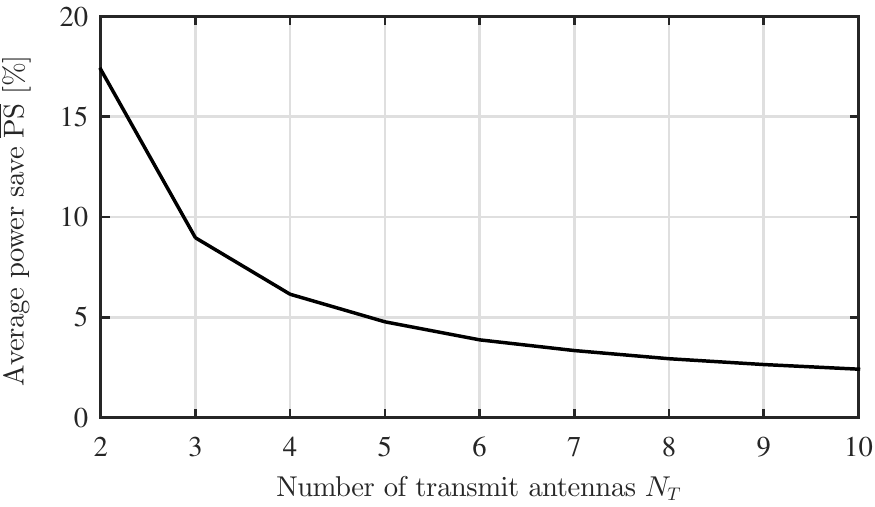} 
\caption{Average throughput gain and average power saving with respect to applying ZF to the SI, with $c=-110$~dB and for different numbers of transmit antennas $N_{T}$.} \label{fig:gain1}
\end{figure}

\begin{figure}[t!]
	\centering
	\includegraphics[scale=1]{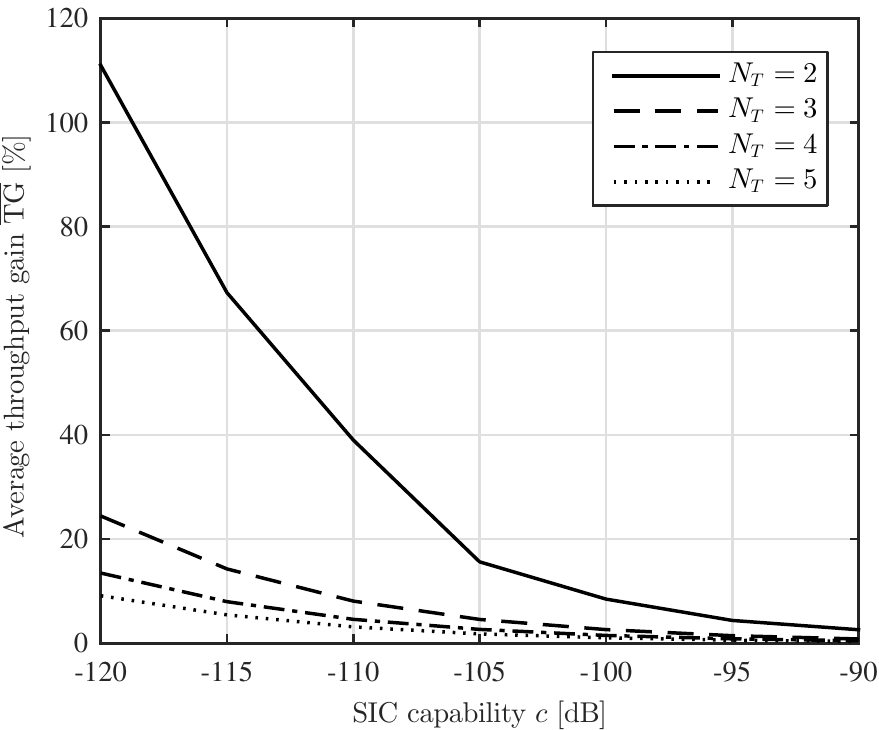} \\ \vspace{2mm}
	\includegraphics[scale=1]{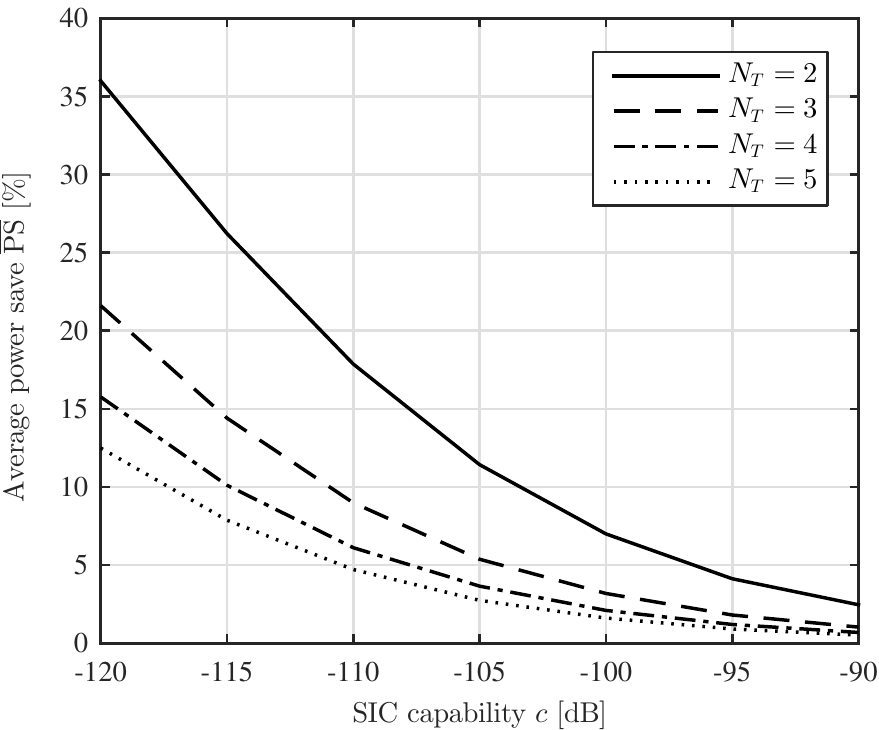}
	\caption{Average throughput gain and average power saving with respect to applying ZF to the SI, with $\rho=0$~dB and for different values of the SIC capability at the receive chain $c$.} \label{fig:gain2}
\end{figure}

\vspace{2mm}

\noindent{\textbf{Performance gains.}} We compare now our optimal transmit beamforming design and the ZF approach presented in \cite{Rii11}, whose complexities can be considered the same as a first approximation. The performance gain of the proposed method in this case can be seen from two perspectives:
\begin{enumerate}
	\item \textit{Average throughput gain}, i.e., the increase in terms of throughput that is obtained for identical transmit power, defined as
	\begin{align*}
	\overline{\mathrm{TG}} & \triangleq \Exp \bigg[ \frac{\log_{2} \big( 1 + \rho |\h_{\mathrm{d}}^{\herm} \w^{\star}|^{2} \big)}{\log_{2} \big( 1 + \rho |\h_{\mathrm{d}}^{\herm} \w_{\mathrm{ZF}}|^{2} \big)} \bigg] - 1;
	\end{align*}
	\item \textit{Average power saving}, i.e., the power reduction that can be supported by the FD BS assuming a target throughput as the one achieved by the ZF approach in \cite{Rii11}, defined as
	\begin{align*}
	\overline{\mathrm{PS}} & \triangleq 1 - \Exp \bigg[ \frac{|\h_{\mathrm{d}}^{\herm} \w_{\mathrm{ZF}}|^{2}}{|\h_{\mathrm{d}}^{\herm} \w^{\star}|^{2}} \bigg].
	\end{align*}
In this regard, we note that $\overline{\mathrm{PS}}$ is independent of the SNR, as intuitively should be, confirming its consistency.  
\end{enumerate}
These metrics are depicted in Figure~\ref{fig:gain1} for $N_T \in [2,10]$ and SNR $\rho \in [-10, 20]$~dB. We start by focusing on $\overline{\mathrm{TG}}$ and observe that, for $c=-110$~dB and the simplest BS setup with $N_{T}=2$ transmit antennas, we obtain an average throughput gain of $29.66\%$ and $11.1\%$ for $\rho=-10$~dB and $\rho=20$~dB, respectively. Two observations are in order at this stage. First, $\overline{\mathrm{TG}}$ decreases as $N_T$ increases: this effect is expected and is due to the impact of the loss of one degree of freedom over the increasing number of available ones (i.e., one transmit antenna is sacrificed for nulling the SI) that characterizes the ZF approach. In this sense, the proposed transmit beamforming design proves particularly suitable for FD radios equipped with small to moderate number of antennas, e.g. FD small-cell BSs. Second, $\overline{\mathrm{TG}}$ decreases as the SNR increases: this is intuitively due to the different impact that the same power gain at the receiver has on the spectral efficiency of the link for different SNR values. In other words, the power gain induced by the proposed solution over \cite{Rii11} results in lower spectral efficiency gains as the SNR increases. Switching our focus on $\overline{\mathrm{PS}}$, we observe that the smaller $N_T$, the larger the average power saving, e.g. $\overline{\mathrm{PS}}=17.87\%$ for $N_T=2$. As for $\overline{\mathrm{TG}}$, the performance gains brought by our optimal transmit beamforming design with respect to applying ZF to the SI decrease as the number of transmit antennas at the FD BS increases, for the same aforementioned reasons. 

Lastly, we study the impact of the SI threshold $\varepsilon$ on both $\overline{\mathrm{TG}}$ and $\overline{\mathrm{PS}}$, by computing these metrics in Figure~\ref{fig:gain2} for different pre-existing SIC capabilities of the FD device, i.e., $c \in [-120, -90]$~dB (see \eqref{eq:epsilon}), and SNR $\rho=0$~dB. We first observe that the proposed technique is more beneficial in terms of $\overline{\mathrm{TG}}$  as the pre-existing SIC capabilities increase, and allows $\overline{\mathrm{TG}}$  to range up to $110.21\%$. This result is rather intuitive to explain, given that the larger the pre-existing SIC capabilities, the more the degrees of freedom loss (due to using the legacy ZF-based method) affects the achievable downlink throughput. The same holds for $\overline{\mathrm{PS}}$, as previously discussed. In practice, and as it could have been expected, more sophisticated SIC strategies allow for larger power saving in terms of transmit power of the FD device. Finally, we note that larger values of $\overline{\mathrm{PS}}$, i.e., up to a remarkable $36.12\%$, are achievable for small values of $N_T$, confirming the findings in Figure~\ref{fig:gain1}.

\vspace{2mm}

\section{Conclusions} \label{sec:Concl}

In this paper, we consider full-duplex (FD) radios with multiple antennas and analyze the problem of identifying the optimal transmit beamforming that maximizes the downlink throughput, while fulfilling the self-interference (SI) cancellation requirements imposed by the receive chain. In this context, the FD radio is subject to strong limitations in terms of transmit power, which cannot exceed a certain threshold in order to protect the incoming signal from the SI. Given the current state of the art solutions, this problem has particular relevance for outdoor small cells populated by mobile users with rapid variations of their fast fading component and for vehicular small cells. In this regard, we derive a closed-form expression for the optimal solution to the considered problem. Remarkably, our numerical findings confirm that the proposed method improves the state of the art in terms of downlink throughput (with respect to applying zero-forcing to the SI) or complexity (with respect to existing solutions based on iterative algorithms). Quantitatively, the magnitude of the achievable gains depends on the pre-existing SI cancellation capabilities and the number of transmit antennas at the FD radio. In particular, both the performance enhancement and power saving grow as the number of antennas decreases and as the SI cancellation at the receive chain increases.

\vspace{2mm}

\balance
\addcontentsline{toc}{chapter}{References}
\bibliographystyle{IEEEtran}
\bibliography{IEEEabrv,ref_Huawei}

\end{document}